# Electronic Landscape of Kagome Superconductors $A$V$_3$Sb$_5$ ($A$ = K, Rb, Cs) from Angle-Resolved Photoemission Spectroscopy

Yong Hu[1,*], Xianxin Wu[2,*], Andreas P. Schnyder[3,*], and Ming Shi[1,4,*]

[1]Photon Science Division, Paul Scherrer Institut, CH-5232 Villigen PSI, Switzerland
[2]CAS Key Laboratory of Theoretical Physics, Institute of Theoretical Physics, Chinese Academy of Sciences, Beijing 100190, China
[3]Max-Planck-Institut für Festkörperforschung, Heisenbergstrasse 1, D-70569 Stuttgart, Germany
[4]Center for Correlated Matter and Department of Physics, Zhejiang University, Hangzhou 310058, China

*To whom correspondence should be addressed:
Y.H. (yonghphysics@gmail.com); X.W. (xxwu@itp.ac.cn); a.schnyder@fkf.mpg.de; M.S. (ming.shi@psi.ch)

**The recently discovered layered kagome superconductors $A$V$_3$Sb$_5$ ($A$ = K, Rb, Cs) have garnered significant attention, as they exhibit an intriguing combination of superconductivity, charge density wave (CDW) order, and nontrivial band topology. As such, these kagome systems serve as an exceptional quantum platform for investigating the intricate interplay between electron correlation effects, geometric frustration, and topological electronic structure. A comprehensive understanding of the underlying electronic structure is crucial for unveiling the nature and origin of the CDW order, as well as determining the electron pairing symmetry in the kagome superconductors. In this review, we present a concise survey of the electronic properties of $A$V$_3$Sb$_5$, with a particular focus on the insights derived from angle-resolved photoemission spectroscopy (ARPES). Through the lens of ARPES, we shed light on the electronic characteristics of the kagome superconductors $A$V$_3$Sb$_5$, which will pave the way for exciting new research frontiers in kagome-related physics.**

## Introduction

The investigation of quantum physics intertwined with nontrivial lattice geometries and strong electronic interactions has emerged as a new frontier in condensed-matter physics. Transition-metal based kagome materials, owing to their unique correlation effects and frustrated lattice geometry inherent to kagome lattices, offer a versatile platform for investigating unconventional correlated and topological quantum states. Depending on the degree of electron filling and interaction within the kagome lattice, a wide range of electronic instabilities has been predicted, including quantum spin liquid [1-3], unconventional superconductivity [4-8], charge density wave (CDW) orders [6-8], and Dirac/Weyl semimetals [9-11]. Particularly noteworthy are the recently discovered non-magnetic vanadium-based superconductors $A$V$_3$Sb$_5$ ($A$ = K, Rb, Cs) (Fig. 1a-d), which have attracted significant attention due to their fascinating electronic phenomena [12-14]. The combination of unconventional CDW, superconductivity, lattice frustration, and topology has sparked rapid exploration in this class of materials, leading to the discovery and investigation of novel electronic phenomena and properties (Fig. 1e), such as three-dimensional (3D) CDW order [15,16], pair density wave (PDW) [17], electronic nematicity [18,19], double superconducting domes under pressure [20-22] and giant anomalous Hall effect [23]. Furthermore, the 3D CDW order, featuring an in-plane 2 × 2 reconstruction, exhibits exotic properties including time-reversal symmetry breaking (TRS) [19,24-26], rotational symmetry breaking [17,18,27] (Fig. 1f), and an intriguing intertwining with unconventional superconductivity [26,28].

In this paper, we review the recent advancements in understanding the electronic structure, CDW order, and superconductivity of the kagome superconductors $A$V$_3$Sb$_5$, derived from angle-resolved photoemission spectroscopy (ARPES). This paper is organized as follows. First, we discuss the crystal structure and band topology of $A$V$_3$Sb$_5$ from the perspective of density functional theory (DFT). Second, we review both the theoretical understanding and experimental evidence of the rich nature of van Hove singularities (VHSs), key ingredients in the electronic structure of the kagome superconductors. Third, we present the spectroscopic fingerprints of the CDW in the electronic structure, including Fermi surface and band reconstructions in the CDW phase. Additionally, we discuss the nature and origin of the CDW order. Fourth, we provide an update on the current understanding of the superconducting gaps of $A$V$_3$Sb$_5$. Finally, we conclude by highlighting the remaining issues to be investigated and provide a future perspective on this research field.

**Crystal structure and band topology**

$A$V$_3$Sb$_5$ in its pristine phase crystalizes in a layered structure with the space group P6/mmm (No. 191) (Fig. 1a). It consists of alternating layers of V-Sb sheets and alkali metal layers. Each V-Sb sheet comprises a two-dimensional (2D) vanadium kagome net intertwined with a hexagonal lattice of Sb atoms (Fig. 1b). Figure 1g displays the bulk Brillouin zone (BZ) and the projected 2D surface BZ. Examination of the phonon band structure calculated for CsV$_3$Sb$_5$, as depicted in Fig. 1h, reveals the softening of acoustic phonon modes near the BZ boundary, specifically around the $M$ and $L$ points, indicating siginificant structural instabilities [29,30]. Consequently, two types of distortions arising from breathing phonon modes are proposed as potential candidates for the CDW structures with a 2 × 2 supercell: a star of David (SoD) (Fig. 1c) and its inverse structure (trihexagonal, TrH) (Fig. 1d).

Investigations of the electronic structure of pristine CsV$_3$Sb$_5$ along high-symmetry directions in the BZ (Fig. 1g), based on DFT calculations, reveals a diverse non-trivial band topology (Fig. 1j). The presence of a direct gap (orange/blue shaded area in Fig. 1j) carrying a non-trivial $\mathbb{Z}_2$ topological index suggests the emergence of topological surface states (TSSs) within the local bandgap at the $M$ point [12,31]. Furthermore, the DFT band dispersions display characteristic features expected from the frustrated kagome lattice (Fig. 1k), including a Dirac cone (DC) at the $K$ point (highlighted by the red arrow in Fig. 1j) [12], a flat band [30] and VHSs around the $M$ point (indicated by the black arrows and labeled as VHS1–4). Spectroscopic evidence supporting the existence of the TSSs, the characteristic Dirac cone, and flat band in CsV$_3$Sb$_5$ has been obtained through ARPES measurements [12,31]. Additionally, the DFT calculations reveal the presence of multiple VHSs in the vicinity of the Fermi level ($E_F$), which have been recognized as playing a crucial role in the intriguing correlated phenomena of the kagome metals $AV_3Sb_5$ [32-36].

**Rich nature of VHSs**

The kagome lattice hosts three distinct sublattices, denoted as A, B, and C (labeled in Fig. 2a). Accordingly, the band structure of the kagome lattice exhibits two types of VHSs: sublattice-pure (*p*-type) and sublattice-mixed (*m*-type) [6,7,35]. In the *p*-type VHS, the Bloch states near the three *M* points originate from mutually different sublattices (Fig. 2b). On the other hand, the *m*-type VHS is characterized by eigenstates that are evenly distributed over two of the three sublattices for each *M* point (Fig. 2c). With respect to the band dispersion around the saddle point, VHSs can be classified into two categories: conventional and higher-order, as illustrated in Fig. 2d and e [37-39]. The conventional VHS displays a saddle-like dispersion (Fig. 2d), leading to a logarithmical divergent density of states (DOS) in 2D. In contrast, the higher-order VHS exhibits a flat dispersion along one direction (Fig. 2e), giving rise to a power-law divergent DOS. In the DFT band structure of $CsV_3Sb_5$, VHS1 exhibits a pronounced flatter dispersion along the *MK* direction compared to the orthogonal direction (Fig. 1j), indicating its higher-order nature [38,39]. This distinguishes it from the other three VHSs near $E_F$, which are all of the conventional type.

Consistent with the theoretical understanding, ARPES measurements have provided compelling evidence, clearly identifying the VHSs and confirming their corresponding sublattice characters (Fig. 2f-n). As shown in Fig. 2h and i, the saddle point nature of the VHS2 and VHS1 is clearly observed through vertically taken cuts across the $\overline{K}\overline{M}$ path (Cut 1-6, as indicated in Fig. 2g), where the band bottom of the electron-like band (solid curve in Fig. 2h and i) exhibits a maximum energy very close to the $E_F$ at the $\overline{M}$ point (dashed curve). Additionally, the hole-like bands (highlighted by the solid curves in Fig. 2j) have a minimum energy at the $\overline{M}$ point, indicating their electron-like nature along the orthogonal direction. These distinct features illustrate the presence of two VHSs above and below the DC, referred to as VHS4 and VHS3, respectively. Notably, VHS1 stands out with a pronounced flat dispersion that extends over more than half of the $\overline{K}\overline{M}$ path (Fig. 2k). The fitting of the experimental spectrum reveals that the quadratic term is substantially smaller than the quartic one (Fig. 2l), highlighting the higher-order nature of VHS1 [40,41].

The sublattice nature of the four VHSs was determined through polarization-dependent ARPES measurements (Fig. 2f, m and n) [40]. Based on the selection rules in photoemission, the orbital characters of the bands forming the VHSs below $E_F$ were identified, as summarized in Fig. 2k. Specifically, VHS1, VHS2, and VHS3 were assigned to the $d_{x^2-y^2}/d_{z^2}$, $d_{yz}$ and $d_{xy}$ orbitals, respectively. These experimental findings align remarkably well with the orbital-resolved band dispersion calculated for the one mirror-invariant sublattice [40]. Moreover, theoretical calculations reveal that the Bloch states associated with VHS1, VHS2 and VHS3 near the $\overline{M}$ point, characterized by $A_g$, $B_{2g}$ and $B_{1g}$ irreducible representations (irrep.), exclusively originate from this sublattice and possess inversion-even parity, thereby confirming their *p*-type nature. In contrast, the VHS4 band corresponds to the $B_{1u}$ irrep. (inversion-odd), and VHS4 originates from the other two sublattices, indicating its *m*-type nature [35, 40,41].

**Band reconstructions in the CDW phase**

The CDW in $A$V$_3$Sb$_5$ demonstrates intriguing properties, such as TRS breaking [19,24-26] and rotational symmetry breaking (Fig. 1f)[17,18,27]. The origin of the CDW order has been theoretically attributed to two scenarios: phonon softening (Fig. 1g-i) [29,30] and correlation-driven Fermi surface instability, *i.e.*, a loop current order (Fig. 1k-m) [32-36]. Despite extensive studies, the mechanism of the CDW formation remains controversial. The presence of a softening acoustic phonon mode around the *L* point (Fig. 1h) implies a 3D nature of the CDW order in $A$V$_3$Sb$_5$. This finding is substantiated by ARPES measurements [42-44], as well as x-ray diffraction (XRD) [15,45-47] and scanning tunneling microscopy (STM) experiments [16], which provide evidence for either a 2 × 2 × 2 or a 2 × 2 × 4 lattice modulation. The spectroscopic fingerprints of the CDW, specifically the reconstructions of the electronic structure observed through ARPES [42-44,48-50], play a crucial role in understanding the nature and distortion pattern of the CDW order, as well as its interplay with superconductivity.

**In-plane band reconstruction**. With multiple VHSs located around the $\bar{M}$ point (Fig. 1j, and Fig. 2), the 3D CDW in $A$V$_3$Sb$_5$, featuring an in-plane 2 × 2 modulation, can induce prominent in-plane electronic reconstructions (Fig. 3). The in-plane component of the CDW results in the folding of the pristine BZ into a smaller BZ (Fig. 3a). The CDW-induced band folding is evident in the measured band structure depicted in Fig. 3b, where an electron-like band around the $\bar{\Gamma}$ exhibits similarities to the one observed at the $\bar{M}$ point [48]. Furthermore, the momentum distribution curve (MDC) (Fig. 3c) near $E_F$ and energy distribution curves (EDCs) (Fig. 3d) around the $\bar{M}$ and $\bar{\Gamma}$ points indicate that the electron-like bands at $\bar{\Gamma}$ and $\bar{M}$ align with the CDW wavevector Q$_1$ (Fig. 3a and c) and share a similar lineshape near $E_F$. Additionally, a band gap associated with the in-plane CDW modulation is identified along the $\bar{\Gamma}$-$\bar{K}$ direction (Fig. 3e) [42,48], which is well reproduced by the DFT calculation considering the 2 × 2 TrH reconstruction (Fig. 3f).

Figures 3g and 3h compare the constant energy contours (CECs) obtained at temperatures well above (200 K, Fig. 3g) and below (20 K, Fig. 3h) the CDW transition temperature ($T_{\text{CDW}}$, 94 K in CsV$_3$Sb$_5$), revealing notable band reconstructions in both the circular-shaped pocket near the zone center ($\bar{\Gamma}$ point) contributed by the Sb *p*-orbitals and the vanadium *d*-orbital bands. Specifically, the electron pocket around $\bar{\Gamma}$ undergoes doubling, while the triangle-shaped CECs around $\bar{K}$ expand in the CDW phase. The band dispersions along the $\bar{\Gamma}$-$\bar{K}$ and $\bar{\Gamma}$-$\bar{M}$ directions, presented in Fig. 3i and j, demonstrate the contribution of the energy shift of the VHS bands to the observed band reconstructions [42]. Moreover, distinctive anisotropic CDW gaps are observed for the Fermi surface derived from the vanadium 3*d*-orbitals (Fig. 3k) [48].

**Out-of-plane band reconstruction.** The out-of-plane component of the 3D CDW also folds the BZ along the $k_z$ direction, leading to an out-of-plane band folding, as illustrated in Fig. 4a. ARPES measurements confirm the presence of this out-of-plane band reconstruction, revealing the characteristic double-band splitting (indicated by the red dashed line and arrow in Fig. 4b and c) in the CDW state. To gain deeper insights into the microscopic structure and properties of the 3D CDW, Figure 4d presents band structures within a 2 × 2 × 2 CDW configuration obtained from DFT calculations, considering four possible configurations: SoD-π, TrH-π, SoD-TrH-π, and SoD-TrH. Notably, the calculations incorporating the superimposed SoD and TrH CDW patterns, namely SoD-TrH-π and SoD-TrH (Fig. 4d), remarkably reproduce the observed double-band splitting features along the $\overline{\Gamma}$-$\overline{K}$ direction. These consistent findings from both ARPES measurements and theoretical calculations demonstrate the intrinsic coexistence of SoD and TrH patterns within the CDW order [42,43], emphasizing the 3D nature of the CDW in the CsV$_3$Sb$_5$.

Moreover, doping-dependent measurements indicate a transition from the alternating SoD- and TrH-like distortions to TrH-π pattern [43], as evidenced by the disappearance of the double-band splitting along the $\overline{\Gamma}$-$\overline{K}$ (Fig. 4e). Intriguingly, this structural transition of the 3D CDW in CsV$_3$Sb$_{5-x}$Sn$_x$ occurs near the maximum of the first superconducting dome (Fig. 4f), implying that the transition to the TrH-π phase and its competition with superconductivity might contribute to the suppression of $T_c$ at intermediate Sn concentrations [43]. In contrast, this CDW pattern transition appears to be absent in KV$_3$Sb$_{5-x}$Sn$_x$, which provides a possible explanation for the lower superconducting transition temperature ($T_c$ = 0.92 K) and the single superconducting dome as a function of Sn substitution (Fig. 4g). Taken together, these doping-dependent results (Fig. 4f and g) offer insights into the emergence of a double-superconducting dome in the phase diagram of CsV$_3$Sb$_5$ [20-22,43,51].

It's worth noting that the band reconstructions observed by ARPES [42,43] may be influenced by factors, such as sample preparation or specific termination conditions. It seems that these band reconstructions are more readily observed in Cs-rich samples or on Cs-terminated surfaces, while they appear to be absent on Sb-terminated surfaces [44,52-54]. Nevertheless, the coexistence of SoD and TrH patterns in the CDW, as deduced from ARPES measurements, aligns consistently with the findings from XRD [45,55], nuclear quadrupole resonance (NQR) [56], and nuclear magnetic resonance (NMR) [57] measurements.

**Superconducting gap**

$A$V$_3$Sb$_5$ exhibits diverse correlated electronic phenomena, including CDW, superconductivity, PDW, and electronic nematicity, reminiscent of the complex competing orders observed in high-temperature superconductors [58]. The fascinating coexistence of these unconventional orders in $A$V$_3$Sb$_5$ highlights the importance of investigating superconducting pairing mechanisms within the realm of kagome superconductors.

**Nodeless superconductivity.** Determining the superconducting gap symmetry is crucial for understanding the underlying mechanism of superconductivity. In CsV$_3$Sb$_5$, evidence such as the presence of a Hebel-Slichter coherence peak just below $T_c$ in ultra-low field NQR measurement and a decrease in the Knight shift below $T_c$, suggests an $s$-wave spin-singlet superconductor [59].

Furthermore, the exponential temperature-dependent behavior of the magnetic penetration depth at low temperatures indicates an absence of nodal quasiparticles [60,61]. However, observations of V-shaped gaps and relatively large residual Fermi-level states in STM measurements [17,62], along with a finite residual thermal conductivity towards zero temperature [63], point to a nodal superconducting gap.

ARPES is a powerful technique for directly measuring the momentum-space structure of the superconducting gap [64]. However, its application in pristine $CsV_3Sb_5$ is challenging due to the relatively low $T_c$ (2.5 K) and small gap size [16,17,2]. Nonetheless, by substituting V with Nb/Ta, the $T_c$ of $CsV_3Sb_5$ can be effectively enhanced to above 4 K [65,66,67]. Utilizing ultrahigh-resolution and low-temperature laser-ARPES, superconducting coherence peaks (Fig. 5a) are observed in EDCs at the Fermi momentum ($k_F$) of the α, β, and δ Fermi surfaces (Fig. 5b) in $Cs(V_{0.93}Nb_{0.07})_3Sb_5$ at 2 K. The extracted superconducting gaps from these EDCs, obtained by fitting to a BCS spectral function, show a nearly isotropic SC gap structure in the Nd-doped sample (Fig. 5c and d). Similar isotropic and orbital-independent superconducting gaps are also observed in $Cs(V_{0.86}Ta_{0.14})_3Sb_5$ (Fig. 5e). The Nb-doped sample exhibits a $T_c$ of 4.4 K and a $T_{CDW}$ of 58 K, while the Ta-doped sample has a higher $T_c$ of 5.2 K without a CDW transition. Regardless of the presence or absence of the CDW, both samples exhibit isotropic gap structures, suggesting a robust nodeless pairing in $CsV_3Sb_5$-derived kagome superconductors [66,68].

**Electron-phonon coupling.** The presence of robust isotropic superconducting gaps seems to be consistent with a conventional *s*-wave pairing mechanism [69]. This is further supported by the observed band dispersion kinks (Fig. 5f-k) [48,70,71], which arise from electron-phonon coupling (EPC). Specifically, both the α band (contributed by Sb-*p* orbitals) and the β band (formed by V-*d* orbitals) exhibit intensity and dispersion anomalies (Fig. 5f-h). The effective real part of the electron self-energy (Re Σ) shows a prominent peak at ~32 meV on both the α and β bands (Fig. 5i), confirming the existence of the dispersion kink. Additionally, a 12-meV kink is specifically observed on the β band (Fig. 5h and i). Interestingly, the EPC strength (λ) associated with the 32-meV kink on the α and β bands displays a nearly isotropic behavior [70]. The experimentally extracted λ in $CsV_3Sb_5$ falls within the intermediate range of 0.45-0.6 (Fig. 5j), which is approximately twice as large as the value predicted by DFT ($λ_{DFT}$ ~0.25) [29]. Moreover, the EPC on the β band is enhanced to λ ~ 0.75 in the isovalent-substituted $Cs(V_{0.93}Nb_{0.07})_3Sb_5$ with an elevated $T_c$ (Fig. 5k) [70].

After discussing the electronic properties of the recently discovered kagome superconductors $AV_3Sb_5$, including the band topology, the nature of VHSs, the band reconstructions in the CDW phase, and the superconducting gap, we now turn our attention to a number of open issues in our understanding of the kagome systems.

**Origin of CDW and TRS breaking.** Two possible scenarios, namely correlation-driven Fermi surface instability (Fig. 1k-m) [32-36] and phonon softening (Fig. 1h and i) [29,30], have been proposed to account for the CDW order in $CsV_3Sb_5$. Both scenarios find support from ARPES experiments, where multiple VHSs in the vicinity to $E_F$ have been observed (Fig. 2). The conventional p-type VHS2 bands feature a good Fermi surface nesting (Fig. 3b-d), with the nesting vector connecting parts of the Fermi surfaces dominated by different sublattices, which can lead to a 2 × 2 bond CDW instability. This provides a plausible explanation for the observed CDW order. Additionally, the coexistence of the SoD and TrH patterns in the CDW order, stemming from the double-band splitting (Fig. 4), suggests the involvement of phonons in driving the CDW. However, since phonon-induced CDW alone cannot account for the observed TRS breaking, EPC and electron-electron interactions may conspire (Fig. 1i) to generate the unconventional CDW order in these vanadium-based kagome metals.

**The underlying mechanism of superconductivity.** While ARPES measurements provide evidence for nodeless superconductivity and EPC in CsV$_3$Sb$_5$ (Fig. 5), supporting a conventional s-wave pairing mechanism, these results do not rule out the possibility of other nodeless pairing states due to the lack of phase information. Muon spin relaxation (μSR) measurements suggest the presence of TRS breaking pairing when the CDW order is eliminated [25]. The observed isotropic superconducting gap with TRS breaking seems to be consistent with an (s + is)-wave or (d + id)-wave pairing, which could arise from the unique electronic interactions and EPC associated with the sublattice-dressed VHSs in the kagome lattice. To unravel the precise pairing mechanism for AV$_3$Sb$_5$, further experimental investigations and theoretical analyses are still required.

**Origin of nematicity.** Another intriguing aspect is the origin of the nematicity in AV$_3$Sb$_5$, which is intertwined with density wave order, similar to the case of cuprates. Both NMR and STM measurements offer clear indications of electronic nematicity and its close relationship with the CDW order. Unlike in a tetragonal lattice, AV$_3$Sb$_5$ does not exhibit a distinct nematic splitting in its electronic structure, which poses challenges for direct detection using ARPES measurements alone. Nevertheless, the combination of ARPES measurements with theoretical calculations holds promise in unraveling the nature and underlying mechanism of the nematic order [71]. In a different context, the recent synthesis of a new family of titanium-based kagome superconductors, namely ATi$_3$Bi$_5$ [72], offers an interesting comparison. In contrast to AV$_3$Sb$_5$, the electronic nematicity in ATi$_3$Bi$_5$ occurs in the absence of CDW [73,74]. This provides a unique platform to investigate pure nematicity within the realm of kagome superconductors and its interplay with orbital degrees of freedom [75].

In conclusion, we have illuminated the recent advancements achieved through photoemission experiments, shedding light on the electronic structure of the kagome superconductor AV$_3$Sb$_5$, which exhibits a spectrum of correlated quantum phenomena. Despite their contentious origins, it is evident that electronic correlations wield a dominant influence in driving the intricate behavior witnessed in AV$_3$Sb$_5$, marking it as a unique member among the recently discovered kagome metals. These captivating complexities require further investigations, which may ultimately lead to a deeper understanding of the fascinating kagome systems, and give fresh insights into the frontiers of condensed matter physics.

**DATA AVAILABILITY**

All data needed to evaluate the conclusions in the paper are present in the paper.

**ACKNOWLEDGEMENTS**

The authors would like to thank Binhai Yan, Ronny Thomale, Tilman Schwemmer, Jia-Xin Yin, Xun Shi, Tianping Ying, Yigui Zhong, Sen Zhou, Ruiqing Fu, Xinloong Han, and Peizhe Ding for enlightening discussion. The work was supported by the Swiss National Science Foundation under Grant. No. 200021_188413. Y.H. acknowledges the support from the National Natural Science Foundation of China (Grant No. 12004363). X.W. was supported by the National Natural Science Foundation of China (grant no. 12047503)

**AUTHOR CONTRIBUTIONS**

Y.H. organized the sections and wrote the manuscript with valuable input from X.W.. A.P.S. and M.S. advised and oversaw the overall structure of the manuscript. All authors contributed to the discussion and comment on the paper.

**ADDITIONAL INFORMATION**

**Competing interests:** The authors declare no competing interests.

**REFERENCES**

1. Balents, L. Spin liquids in frustrated magnets. *Nature* **464**, 199–208 (2010).

2. Zhou, Y., Kanoda, K. & Ng, T.-K. Quantum spin liquid states. *Rev. Mod. Phys.* **89**, 025003 (2017).

3. Norman, M.R. Colloquium: herbertsmithite and the search for the quantum spin liquid. *Rev. Mod. Phys.* **88**, 041002 (2016).

4. Ko, W. -H., Lee, P. A. & Wen, X. -G. Doped kagome system as exotic superconductor. *Phys. Rev. B* **79**, 214502 (2009).

5. Yu, S.-L & Li, J.-X. Chiral superconducting phase and chiral spin-density-wave phase in a Hubbard model on the kagome lattice. *Phys. Rev. B* **85**, 144402 (2012).

6. Kiesel, M. L. & Thomale, R. Sublattice interference in the kagome Hubbard model. *Phys. Rev. B* **86**, 121105(R) (2012).

7. Kiesel, M. L., Platt, C. & Thomale, R. Unconventional Fermi surface instabilities in the kagome Hubbard model. *Phys. Rev. Lett.* **110**, 126405 (2013).

8. Wang, W.-S., Li, Z.-Z., Xiang, Y.-Y. & Wang, Q.-H. Competing electronic orders on kagome lattices at van Hove filling. *Phys. Rev. B* **87**, 115135 (2013).

9. Ye, L. *et al.* Massive Dirac fermions in a ferromagnetic kagome metal. *Nature* (London) **555**, 638–642 (2018).

10. Morali, N. *et al.* Fermi-arc diversity on surface terminations of the magnetic Weyl semimetal $Co_3Sn_2S_2$. *Science* **365**, 1286–1291 (2019).

11. Liu, D.-F. *et al.* Magnetic Weyl semimetal phase in a kagomé crystal. *Science* **365**, 1282–1285 (2019).

12. Ortiz, B. R. *et al.* $CsV_3Sb_5$: A $\mathbb{Z}_2$ Topological kagome metal with a superconducting ground state. *Phys. Rev. Lett.* **125**, 247002 (2020).

13. Neupert, T. *et al.* Charge order and superconductivity in kagome materials. *Nat. Phys.* **18**, 137–143 (2022).

14. Jiang, K. *et al.* Kagome superconductors $AV_3Sb_5$ (*A* = K, Rb, Cs). *Natl. Sci. Rev.* **10**, nwac199 (2023).

15. Li, H. X. *et al.* Observation of unconventional charge density wave without acoustic phonon anomaly in kagome superconductors $AV_3Sb_5$ (*A* = Rb, Cs). *Phys. Rev. X* **11**, 031050 (2021).

16. Liang, Z. *et al.* Three-dimensional charge density wave and robust zero-bias conductance peak inside the superconducting vortex core of a kagome superconductor $CsV_3Sb_5$. *Phys. Rev. X* **11**, 031026 (2021).

17. Chen, H. *et al.* Roton pair density wave and unconventional strong-coupling superconductivity in a topological kagome metal. *Nature* **559**, 222–228 (2021).

18. Nie, L. *et al.* Charge-density-wave-driven electronic nematicity in a kagome superconductor. *Nature* **604**, 59–64 (2022).

19. Xu, Y. *et al.* Three-state nematicity and magneto-optical Kerr effect in the charge density waves in kagome superconductors. *Nat. Phys.* **18**, 1470–1475 (2022).

20. Chen, K. Y. *et al.* Double superconducting dome and triple enhancement of $T_c$ in the kagome superconductor $CsV_3Sb_5$ under high pressure. *Phys. Rev. Lett.* **126**, 247001 (2021).

21. Zhang, Z. *et al.* Pressure-induced reemergence of superconductivity in topological Kagome metal $CsV_3Sb_5$. *Phys. Rev. B* **103**, 224513 (2021).

22. Chen, X. *et al.* Highly-robust reentrant superconductivity in $CsV_3Sb_5$ under pressure. *Chin. Phys. Lett.* **38**, 057402 (2021).

23. Yang, S.-Y. *et al.* Giant, unconventional anomalous Hall effect in the metallic frustrated magnet candidate $KV_3Sb_5$. *Sci. Adv.* **6**, abb6003 (2020).

24. Jiang, Y.-X. *et al.* Discovery of unconventional chiral charge order in kagome superconductor $KV_3Sb_5$. *Nat. Mater.* **20**, 1353–1357 (2021).

25. Mielke III, C. *et al.* Time-reversal symmetry-breaking charge order in a correlated kagome superconductor. *Nature* **602**, 245–250 (2022).

26. Guguchia, Z. *et al.* Tunable unconventional kagome superconductivity in charge ordered RbV$_3$Sb$_5$ and KV$_3$Sb$_5$. *Nat. Commun*. **14**, 153 (2023).

27. Zhao, H. *et al.* Cascade of correlated electron states in the kagome superconductor CsV$_3$Sb$_5$. *Nature* **599**, 216–221 (2021).

28. Zheng, L. *et al.* Emergent charge order in pressurized kagome superconductor CsV$_3$Sb$_5$. *Nature* **611**, 682–687 (2022).

29. Tan, H., Liu, Y., Wang, Z. & Yan, B. *et al.* Charge density waves and electronic properties of superconducting kagome metals. *Phys. Rev. Lett.* **127**, 046401 (2021).

30. Christensen, M. H., Birol, T., Andersen, B. M. & Fernandes, R. M. Theory of the charge-density wave in $A$V$_3$Sb$_5$ kagome metals. *Phys. Rev. B* **104**, 214513 (2021).

31. Hu, Y. *et al.* Topological surface states and flat bands in the kagome superconductor CsV$_3$Sb$_5$. *Sci. Bull.* **67**, 495–500 (2022).

32. Lin, Y.-P. & Nandkishore, R. M. Complex charge density waves at van Hove singularity on hexagonal lattices: Haldane-model phase diagram and potential realization in kagome metals $A$V$_3$Sb$_5$. *Phys. Rev. B* **104**, 045122 (2021).

33. Park, T., Ye, M. & Balents, L. Electronic instabilities of kagomé metals: Saddle points and Landau theory. *Phys. Rev. B* **104**, 035142 (2021).

34. Feng, X., Jiang, K., Wang, Z. & Hu, J. Chiral flux phase in the Kagome superconductor $A$V$_3$Sb$_5$. *Sci. Bull.* **66**, 1384 (2021).

35. Wu, X. *et al.* Nature of unconventional pairing in the kagome superconductors $A$V$_3$Sb$_5$. *Phys. Rev. Lett.* **127**, 177001 (2021).

36. Denner, M. M., Thomale, R. & Neupert, T. Analysis of charge order in the kagome metal $A$V$_3$Sb$_5$ ($A$ = K, Rb, Cs). *Phys. Rev. Lett.* **127**, 217601 (2021); **128**, 099901(E) (2022).

37. Van H., Léon. The occurrence of singularities in the elastic frequency distribution of a crystal. *Phys. Rev.* **89**, 1189 (1953).

38. Yuan, N. F. Q., Isobe, H. & Fu, L. Magic of high-order van Hove singularity. *Nat. Commun.* **10**, 5769 (2019).

39. Classen, L., Chubukov, A. V., Honerkamp, C. & Scherer, M. M. Competing orders at higher-order van Hove points. *Phys. Rev. B* **102**, 125141 (2020).

40. Hu, Y. *et al.* Rich nature of van Hove singularities in kagome superconductor CsV$_3$Sb$_5$. *Nat. Commun*. **13**, 2220 (2022).

41. Kang, M. *et al.* Twofold van Hove singularity and origin of charge order in topological kagome superconductor CsV$_3$Sb$_5$. *Nat. Phys*. **18**, 301–308 (2022).

42. Hu, Y. *et al.* Coexistence of trihexagonal and star-of-David pattern in the charge density wave of the kagome superconductor $A$V$_3$Sb$_5$. *Phys. Rev. B* **106**, L241106 (2022).

43. Kang, M. *et al.* Charge order landscape and competition with superconductivity in kagome metals. *Nat. Mater.* **22**, 186–193 (2022).

44. Luo, Y. *et al.* Electronic states dressed by an out-of-plane supermodulation in the quasi-two-dimensional kagome superconductor CsV$_3$Sb$_5$. *Phys. Rev. B* **105**, L241111 (2022).

45. Ortiz, B. R. *et al.* Fermi surface mapping and the nature of charge density wave order in the kagome superconductor CsV$_3$Sb$_5$. *Phys. Rev. X* **11**, 041030 (2021).

46. Stahl, Q. *et al.* Temperature-driven reorganization of electronic order in CsV$_3$Sb$_5$. *Phys. Rev. B* **105**, 195136 (2022).

47. Kautzsch L. *et al*. Structural evolution of the kagome superconductors $AV_3Sb_5$ ($A$ = K, Rb, and Cs) through charge density wave order. *Phys. Rev. Materials* **7**, 024806 (2023).

48. Luo, H. L. *et al*. Electronic nature of charge density wave and electron-phonon coupling in kagome superconductor $KV_3Sb_5$. *Nat. Commun.* **13**, 273 (2022).

49. Liu, Z. *et al*. Charge-density-wave-induced bands renormalization and energy gaps in a kagome superconductor $RbV_3Sb_5$. *Phys. Rev. X* **11**, 041010 (2021).

50. Lou, R. *et al*. Charge-density-wave-induced peak-dip-hump structure and the multiband superconductivity in a kagome superconductor $CsV_3Sb_5$. *Phys. Rev. Lett.* **128**, 036402 (2022).

51. Oey, Y. M. *et al*. Fermi level tuning and double-dome superconductivity in the kagome metal $CsV_3Sb_{5-x}Sn_x$. *Phys. Rev. Materials* **6**, L041801 (2022).

52. Kato, T. *et al*. Polarity-dependent charge density wave in the kagome superconductor $CsV_3Sb_5$. *Phys. Rev. B* **106**, L121112 (2022).

53. Nakayama, K. *et al*. Carrier injection and manipulation of charge-density wave in kagome superconductor $CsV_3Sb_5$. *Phys. Rev. X* **12**, 011001 (2022).

54. Zhu, H. *et al*. Electronic instability of kagome metal $CsV_3Sb_5$ in the 2 × 2 × 2 charge density wave state. *Chin. Phys. Lett.* **40**, 047301 (2023).

55. Xiao, Q. *et al*. Coexistence of multiple stacking charge density waves in kagome superconductor $CsV_3Sb_5$. *Phys. Rev. Research* **5**, L012032 (2023).

56. Feng, X.Y. *et al*. Commensurate-to-incommensurate transition of charge-density-wave order and a possible quantum critical point in pressurized kagome metal $CsV_3Sb_5$. *npj Quantum Mater*. **8**, 23 (2023).

57. Frassineti, J. *et al*. Microscopic nature of the charge-density wave in the kagome superconductor $RbV_3Sb_5$. *Phys. Rev. Research* **5**, L012017 (2023).

58. Keimer, B. *et al*. From quantum matter to high-temperature superconductivity in copper oxides. *Nature* **518**, 179–186 (2015).

59. Mu, C. *et al*. S-wave superconductivity in kagome metal $CsV_3Sb_5$ revealed by $^{121/123}$Sb NQR and $^{51}$V NMR measurements. *Chin. Phys. Lett.* **38**, 077402 (2021).

60. Duan, W. *et al*. Nodeless superconductivity in the kagome metal $CsV_3Sb_5$. *Sci. China Phys. Mech. Astron.* **64**, 107462 (2021).

61. Gupta, R. *et al*. Microscopic evidence for anisotropic multigap superconductivity in the $CsV_3Sb_5$ kagome superconductor. *npj Quantum Mater*. **7**, 49 (2022).

62. Xu, H.-S. *et al*. Multiband superconductivity with sign-preserving order parameter in kagome superconductor $CsV_3Sb_5$. *Phys. Rev. Lett.* **127**, 187004 (2021).

63. Zhao, C. *et al*. Nodal superconductivity and superconducting domes in the topological kagome metal $CsV_3Sb_5$. Preprint at https://doi.org/10.48550/arXiv.2102.08356 (2021).

64. Damascelli, A., Hussain, Z. & Shen, Z.-X. Angle-resolved photoemission studies of the cuprate superconductors. *Rev. Mod. Phys.* **75**, 473–541 (2003).

65. Li, Y. *et al.* Tuning the competition between superconductivity and charge order in the kagome superconductor $Cs(V_{1-x}Nb_x)_3Sb_5$. *Phys. Rev. B* **105**, L180507 (2022).

66. Zhong, Y. *et al.* Nodeless electron pairing in $CsV_3Sb_5$-derived kagome superconductors. *Nature* **617**, 488–492 (2023).

67. Luo, Y. *et al.* A unique van Hove singularity in kagome superconductor $CsV_{3-x}Ta_xSb_5$ with enhanced superconductivity. *Nat. Commun.* **14**, 3819 (2023).

68. Zhang, W. *et al.* Nodeless Superconductivity in Kagome Metal $CsV_3Sb_5$ with and without Time Reversal Symmetry Breaking. *Nano Lett.* **23**, 872–879 (2023).

69. Roppongi, M. *et al.* Bulk evidence of anisotropic *s*-wave pairing with no sign change in the kagome superconductor CsV$_3$Sb$_5$. *Nat. Commun.* **14**, 667 (2023).

70. Zhong, Y. *et al.* Testing electron–phonon coupling for the superconductivity in kagome metal CsV$_3$Sb$_5$. *Nat. Commun.* **14**, 1945 (2023).

71. Jiang, Z. *et al.* Observation of electronic nematicity driven by the three-dimensional charge density wave in kagome lattice KV$_3$Sb$_5$. *Nano Lett.* 23, 5625–5633 (2023).

72. Yang, H. *et al.* Titanium-based kagome superconductor CsTi$_3$Bi$_5$ and topological states. Preprint at https://arxiv.org/abs/2209.03840 (2022).

73. Yang, H. *et al.* Superconductivity and orbital-selective nematic order in a new titanium-based kagome metal CsTi$_3$Bi$_5$. Preprint at https://arxiv.org/abs/2211.12264 (2022).

74. Li, H. *et al.* Electronic nematicity in the absence of charge density waves in a new titanium-based kagome metal. *Nat. Phys.* (2023). https://doi.org/10.1038/s41567-023-02176-3

75. Hu, Y. *et al.* Non-trivial band topology and orbital-selective electronic nematicity in a titanium-based kagome superconductor. *Nat. Phys.* (2023). https://doi.org/10.1038/s41567-023-02215-z

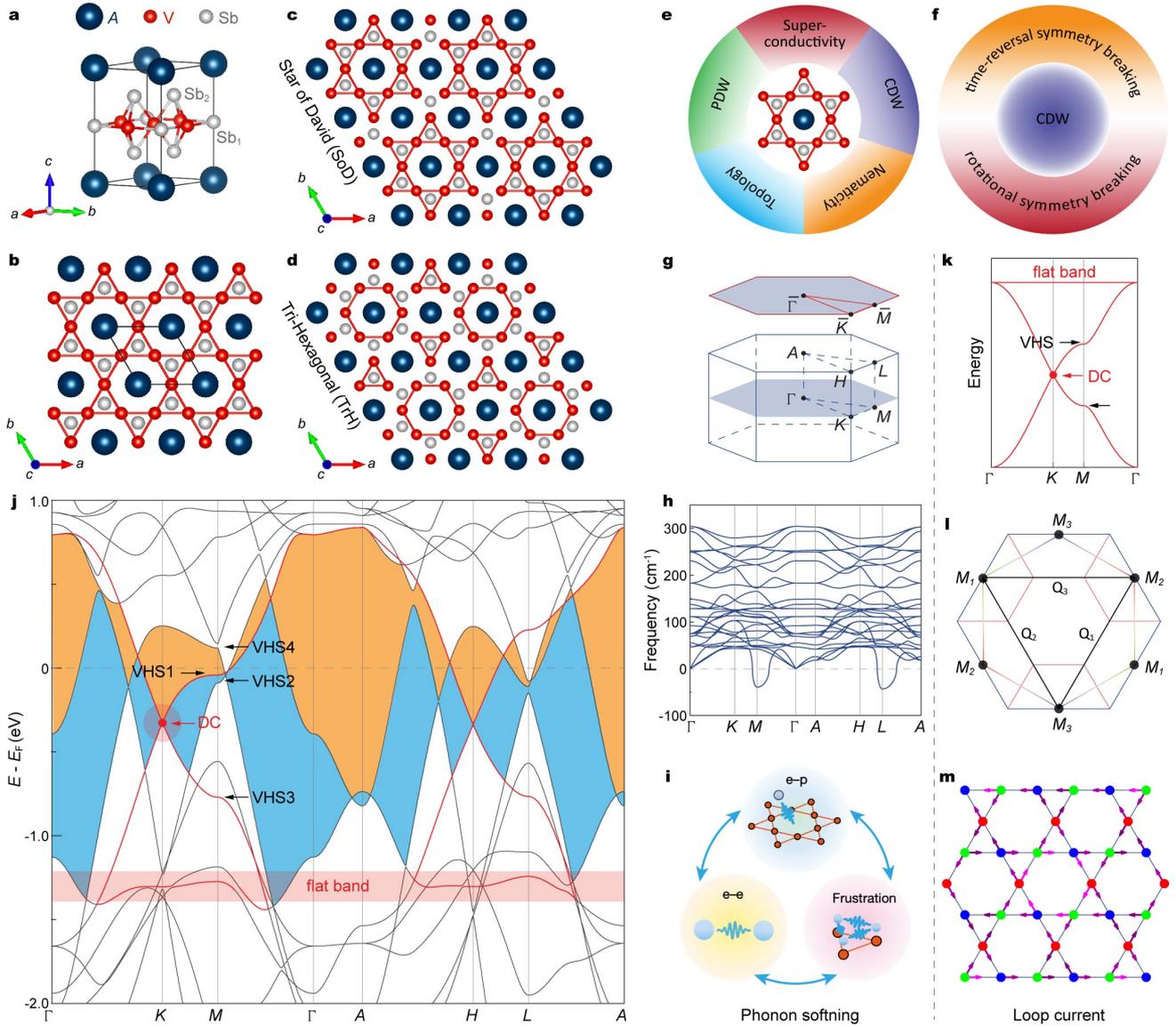

**Fig. 1 | Crystal structure and band topology of $A$V$_3$Sb$_5$. a-d** Crystal structure of $A$V$_3$Sb$_5$ in the normal state [(a) the unit cell; (b) top view showing the kagome plane], and CDW phase with the candidate SoD (c) and TrH distortions (d). **e** The novel electronic phenomena found in CsV$_3$Sb$_5$. **f** Schematic of time-reversal symmetry breaking and rotational symmetry breaking in the CDW phase. **g** Schematic of the bulk and surface Brillouin zones along the (001) surface of CsV$_3$Sb$_5$ with high-symmetry points marked. **h** DFT calculated phonon band structure for pritine CsV$_3$Sb$_5$. **i** Illustration of the interplay among electron-phonon (*e-p*), electron-electron (*e-e*) and geometric frustration in CsV$_3$Sb$_5$ (iii). Adapted from ref. 28. **j** DFT band structure of CsV$_3$Sb$_5$ along high-symmetry directions, with VHS, DC and flat band highlighted. The direct gap (yellow/blue shaded area) carries a non-trivial $\mathbb{Z}_2$ topological index. The 2D kagome dispersion is highlighted in red. **k-m** The nearest-neighbor tight-binding band structure of kagome lattice featuring VHS, DC and flat band (k), nesting wave vectors of a kagome FS at van Hove filling (l), and loop current order originating from VHS (m).

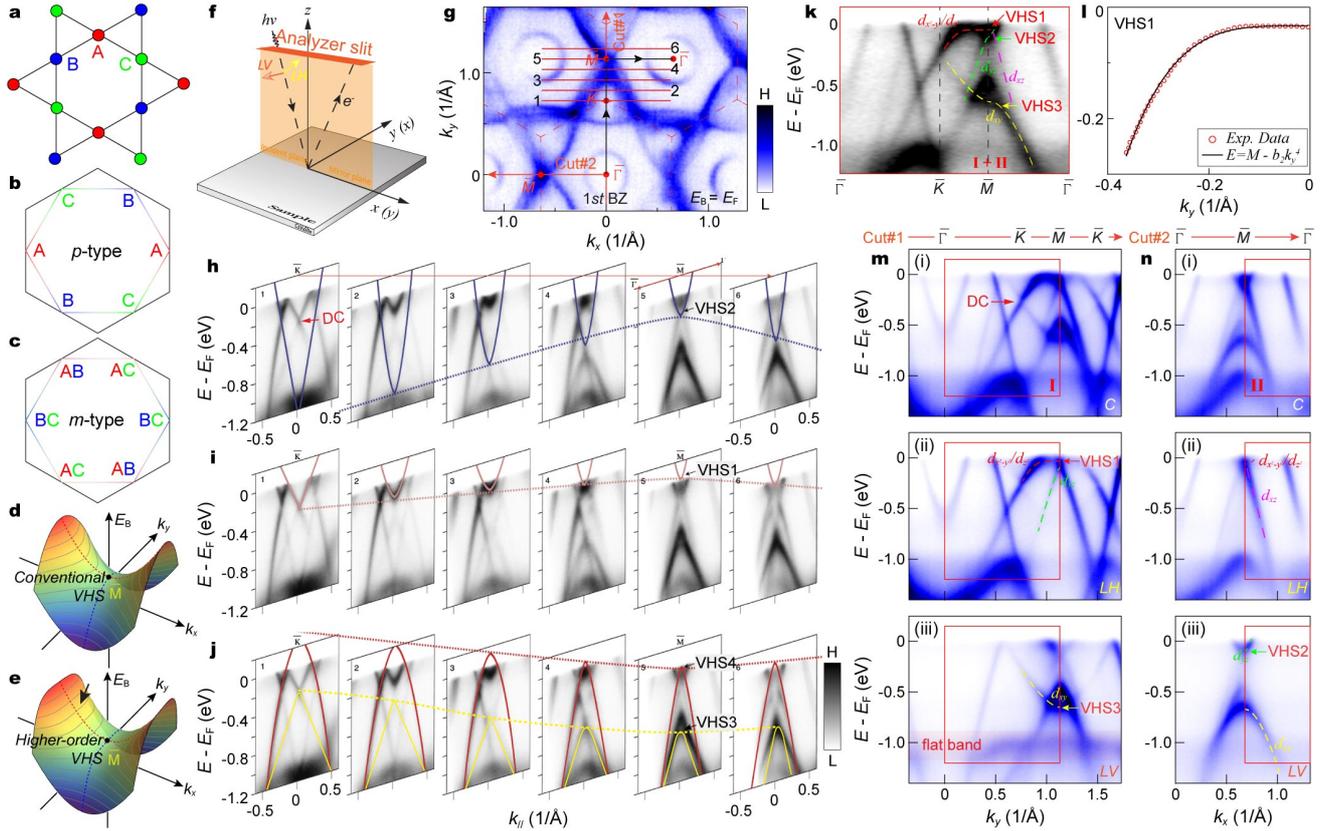

**Fig. 2 | Multiple VHSs and the orbital nature of the kagome bands in CsV$_3$Sb$_5$. a** Real space structure of the kagome vanadium plane, with three sublattices marked. **b,c** Two distinct types of sublattice decorated VHSs, labeled as *p*-type (b) and *m*-type (sublattice mixing) (c). **d,e** Schematics of the conventional VHS (d) and higher-order VHS (e) in 2D electron system. **f** Experimental geometry of the polarization-dependent ARPES. **g** FS of CsV$_3$Sb$_5$. **h-j** A series of cuts taken vertically across the $\overline{K}\overline{M}$ path, the momentum paths of the cuts (1-6) are indicated by the red lines in (**g**). **k** Experimental band dispersion along the $\overline{\Gamma}$-$\overline{K}$-$\overline{M}$-$\overline{\Gamma}$ direction, as indicted by the red arrows in (**g**). **l** Fittings of the measured dispersion along the $\overline{K}\overline{M}$ path. **m** Band dispersion along the $\overline{\Gamma}$-$\overline{K}$-$\overline{M}$ direction [cut#1, indicated by the red arrow in (**g**)], measured with circular (*C*) (i), linear horizontal (*LH*) (ii) and linear vertical (*LV*) (iii) polarizations. **n** Same data as in (**m**), but measured along the $\overline{\Gamma}$-$\overline{M}$ direction [cut#2, indicated by the red arrow in (**g**)]. Experimental band dispersions in (**g-n**) are measured in the normal state. All data are reprinted with permission from Ref. 40, except (**h-j**) which is adapted from Ref. 41.

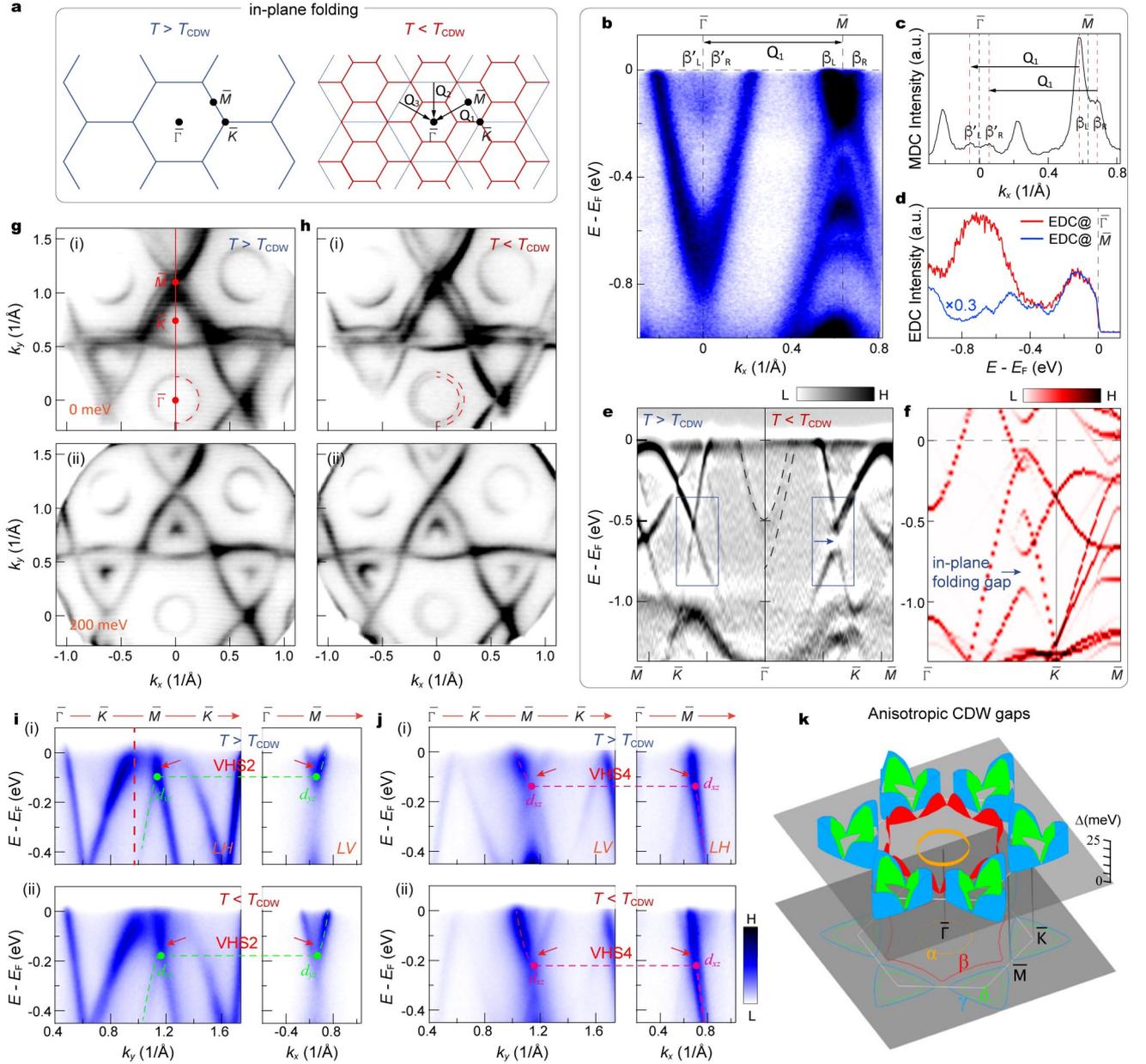

**Fig. 3 | In-plane band reconstruction in $A$V$_3$Sb$_5$. a** Band reconstruction from the in-plane component of CDW indicated by schematics of the in-plane folding of the surface BZ. **b-d** ARPES spectrum of KV$_3$Sb$_5$ measured along the $\overline{\Gamma}$-$\overline{M}$ direction (**b**), the corresponding MDC at $E_F$ (**c**) and EDCs taken at the $\overline{M}$ and $\overline{\Gamma}$ points (**d**). **e** Comparison of the curvature plots of CsV$_3$Sb$_5$ taken along the $\overline{\Gamma}$-$\overline{K}$ direction at 200 K (left) and 20 K (right). **f** DFT the band structure of CsV$_3$Sb$_5$ for the 2D 2 × 2 TrH phase. **g,h** Constant energy contours at the $E_F$ (i) and binding energy of 200 meV (ii) measured at 200 (**g**) and 20 K (**h**). **i** ARPES spectra of CsV$_3$Sb$_5$ taken along the $\overline{\Gamma}$-$\overline{K}$ direction with *LH* (left) and *LV* (right) polarizations, at 200 K (i) and 20 K (ii). **j** Same data as in (**i**), but measured with *LH* polarization along the $\overline{\Gamma}$-$\overline{K}$ direction (left) and *LV* polarization along the $\overline{\Gamma}$-$\overline{M}$ direction (right). **k** Anisotropic CDW gaps in KV$_3$Sb$_5$. Results in (**b-d, k**) are reprinted with permission from ref. 48. Data in (**e-j**) are reprinted with permission from ref. 42 (Copyright 2022 by the American Physical Society).

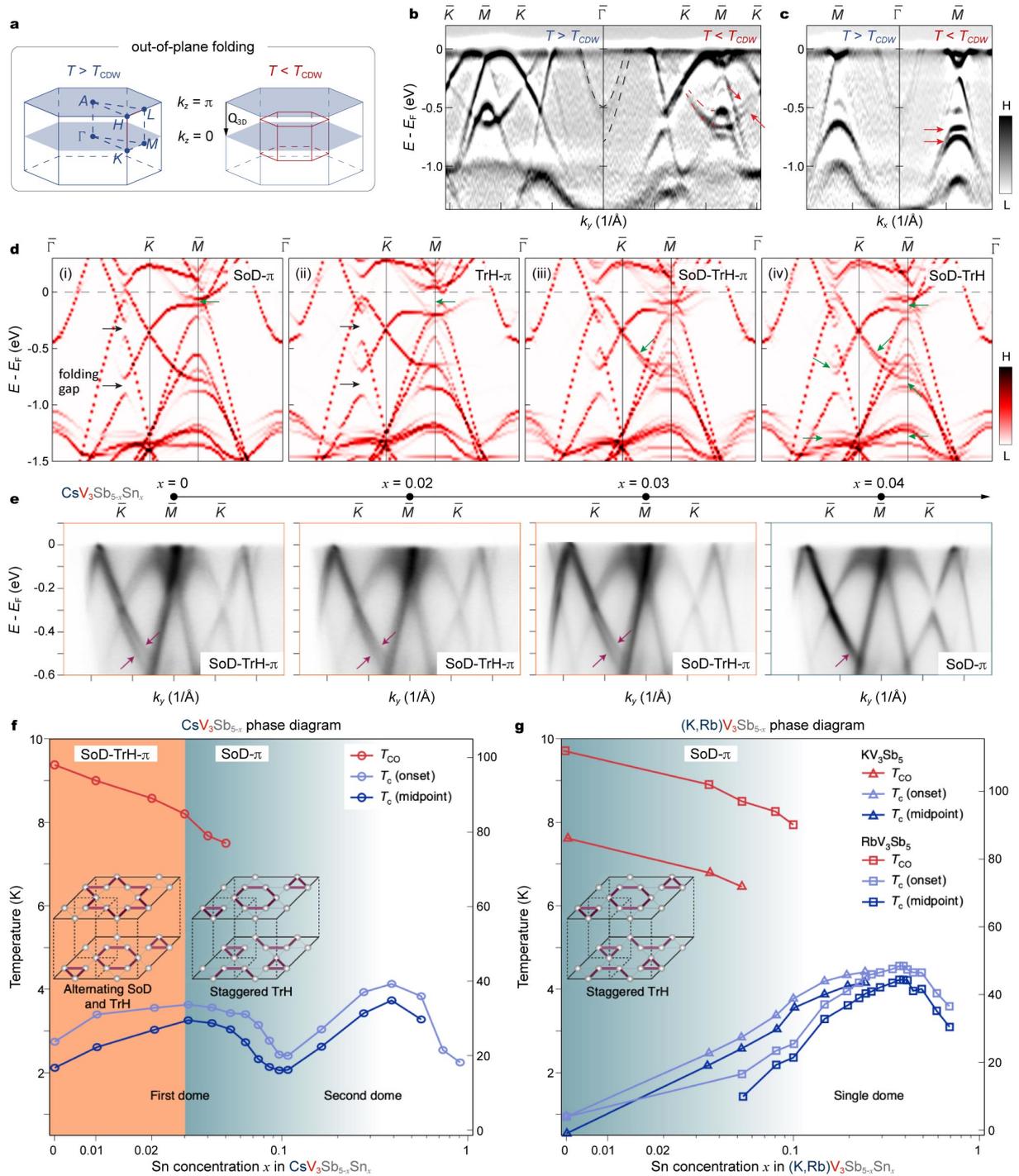

**Fig. 4 | Out-of-plane band reconstruction in $AV_3Sb_{5-x}Sn_x$. a** Band reconstruction from the out-of-plane component of CDW indicated by schematics of the out-of-plane folding of the bulk BZ. **b,c** Comparison of the curvature plots of $CsV_3Sb_5$ taken at 200 K (left) and 20 K (right) along the $\bar{\Gamma}$-$\bar{K}$-$\bar{M}$ direction (**b**) and the $\bar{\Gamma}$-$\bar{M}$ direction (**c**). The double-band splitting is highlighted by dashed curve and arrow. **d** DFT band structures of $CsV_3Sb_5$ for the 3D SoD-π (i), TrH-π (ii), SoD-TrH-π (iii), and SoD-TrH (iv) lattice configurations. **e** Evolution of the band reconstruction in $CsV_3Sb_{5-x}Sn_x$ as a function of Sn doping with x = 0, 0.02, 0.03 and 0.04. The purple arrows highlight the splitting on the Dirac band. **f** Phase diagram of $CsV_3Sb_{5-x}Sn_x$. The open dark blue circles represent the $T_c$. The red circles represent the $T_{CDW}$. The orange and teal backgrounds represent the regions with SoD-TrH-π and SoD-π phases, respectively. **g** Phase diagram of $(K,Rb)Sb_{5-x}Sn_x$ with only the 3D SoD-π CDW phase (inset) and single superconducting dome. Data in (**b**-**d**) and (**e**-**g**) are reprinted with permission from ref. 42 and ref. 43, respectively.

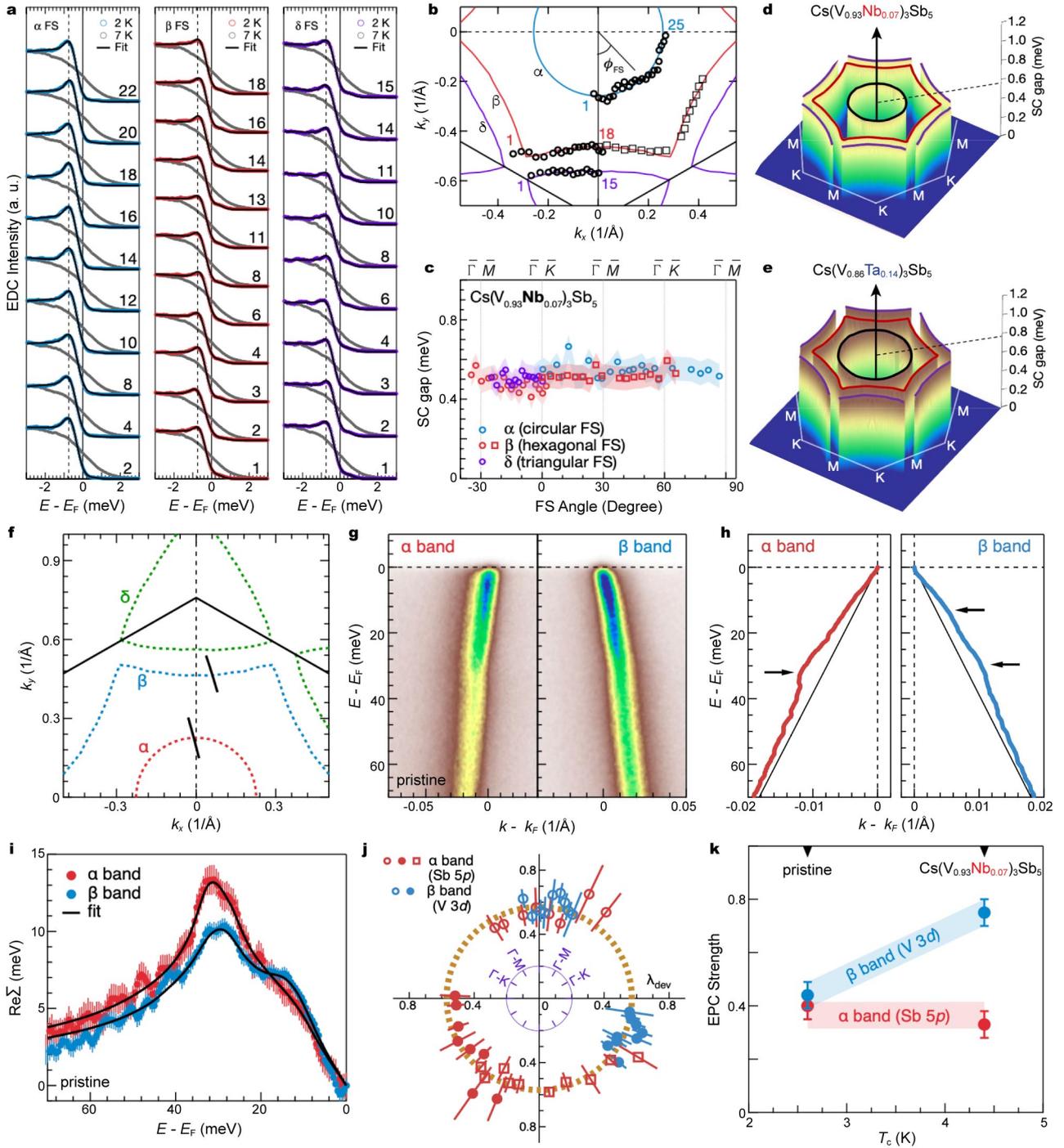

**Fig. 5 | Isotropic superconducting gap and electron-phonon coupling in Cs(V$_{0.93}$Nb$_{0.07}$)$_3$Sb$_5$. a,b** EDCs at $k_F$ taken along the α, β, and δ FS. The positions of the EDCs are indicated in (**b**). The black curves in (**a**) are fitted by the BCS spectral function. The dashed lines in (**a**) mark the estimated peak position of the EDCs. **c** Superconducting gap amplitude estimated from the fits to EDCs shown in (**a**). The square makers are the superconducting gap results from a different sample and the corresponding $k_F$ is indicated by the squares in (**b**). **d,e** Schematic momentum dependence of the superconducting gap amplitude of the Cs(V$_{0.93}$Nb$_{0.07}$)$_3$Sb$_5$ (**d**) and Cs(V$_{0.86}$Ta$_{0.14}$)$_3$Sb$_5$ (**e**) samples, respectively. **f-h** ARPES intensity plots the α and β bands (**g**). The momentum positions of the bands are indicated in (**f**). Extracted band dispersions of the α and β bands (**h**). The black arrows highlight the energy position of the kinks. The black lines are the corresponding bare bands. **i** Extracted ReΣ(ω) of the α and β bands. **j** EPC strength λ defined by the slope of ReΣ(ω) [70]. **k** EPC strength λ as a function of $T_c$ tuned by Nb substitution. The error bar is determined by the standard deviation of the ReΣ(ω) [70]. Data in (**a-e**) and (**f-k**) are reprinted with permission from ref. 66 and ref. 70, respectively.